\documentclass[a4paper,11pt]{article}

\pdfoutput=1 

\usepackage{jcappub} 

\usepackage{graphics}
\usepackage{graphicx,subfigure}
\usepackage[]{algorithm2e}
\usepackage{xspace}
\usepackage{amssymb}
\usepackage{amsmath}
\usepackage{hyperref}
\usepackage{pbox}

\newcommand{\ufig}{\texttt{UFig}\xspace}
\newcommand{\sextractor}{\texttt{Source-Extractor}\xspace}

\title{Approximate Bayesian Computation for Forward Modeling in Cosmology}

\author[a]{Jo\"el Akeret}

\author[a]{Alexandre Refregier}

\author[a]{Adam Amara}

\author[a]{Sebastian Seehars}

\author[a]{Caspar Hasner }

\affiliation[a]{ETH Zurich, Institute for Astronomy, Department of Physics, Wolfgang Pauli Strasse 27, 8093 Zurich, Switzerland}
\emailAdd{joel.akeret@phys.ethz.ch}

\abstract{Bayesian inference is often used in cosmology and astrophysics to derive constraints on model parameters from observations. This approach relies on the ability to compute the likelihood of the data given a choice of model parameters. In many practical situations, the likelihood function may however be unavailable or intractable due to non-gaussian errors, non-linear measurements processes, or complex data formats such as catalogs and maps. In these cases, the simulation of mock data sets can often be made through forward modeling. We discuss how Approximate Bayesian Computation (ABC) can be used in these cases to derive an approximation to the posterior constraints using simulated data sets. This technique relies on the sampling of the parameter set, a distance metric to quantify the difference between the observation and the simulations and summary statistics to compress the information in the data. We first review the principles of ABC and discuss its implementation using a Population Monte-Carlo (PMC) algorithm and the Mahalanobis distance metric. We test the performance of the implementation using a Gaussian toy model. We then apply the ABC technique to the practical case of the calibration of image simulations for wide field cosmological surveys. We find that the ABC analysis is able to provide reliable parameter constraints for this problem and is therefore a promising technique for other applications in cosmology and astrophysics. Our implementation of the ABC PMC method is made available via a public code release.}


\notoc
\begin{document}
\maketitle
\flushbottom

\section{Introduction}
\label{sec:introduction}

Bayesian inference is commonly used in cosmology and astrophysics to derive constraints on the parameters of a model from observations \cite[e.g.][]{Christensen2001, Knox2001}. In this framework, the posterior distribution of the model parameters given the observed data is derived from a prior and from the likelihood of the data given a choice of model parameters. Various sampling techniques are used to constrain high dimensional parameter spaces such as the Metropolis Hastings \cite{Metropolis1953, Hastings1970}, Gibbs sampling \cite{geman1984stochastic}, Nested sampling \cite{skilling2004nested}, Hamiltonian/Hybrid Monte Carlo \cite{duane1987hybrid}, Sequential Monte Carlo \cite{del1996non, cappe2004population} or the more recent affine invariant ensemble sampling method \cite{Goodman2010a, foreman2013emcee}. Various software packages in cosmology have implemented these algorithms and are being used for different applications. Prominent examples are the {\tt cosmomc} package \cite{lewis2002cosmological}, based on Metropolis Hasting, {\tt cosmopmc} \cite{kilbinger2011cosmopmc} which is based on Population Monte Carlo and {\tt CosmoHammer} \cite{akeret2013cosmohammer} that is built on an ensemble sampling algorithm. Popular codes based on Nested sampling are {\tt Multinest} \cite{feroz2009multinest} or {\tt PolyChord} \cite{handley2015polychord}. More recently, packages have appeared that allow the user to switch between different sampling methods, such as {\tt Monte Python}, {\tt Cosmo++} or {\tt Cosmosis}\cite{audren7183conservative, aslanyan2014cosmo, zuntz2014cosmosis}.

Bayesian inference relies on the ability to compute the likelihood function. In practice, there are however situations in which the likelihood function is unknown or computationally intractable and where the direct Bayesian analysis is therefore not possible. In some of these instances, however, the simulation of mock data sets can be made through forward modeling \cite[e.g.][]{kacprzak2012measurement, mesinger201121cmfast, reinecke2006simulation, 2015arXiv150406570P}. The simulations may include a model of the astrophysical signal, the instrument and observing conditions, and of the data analysis pipeline. For example, N-body simulations, semi-analytical models and image simulations can be used to generate simulated galaxy catalogues \cite[e.g.][]{2015ApJ...801...73C}. Another example is the simulation of weak lensing or other extra-galactic maps which are characterised by non-gaussian statistics \cite[e.g.][]{juin2007cosmology, pires2009fast, dietrich2010cosmology, marian2012optimized}, or image simulations for weak lensing measurements \cite[e.g.][]{heymans2006shear, massey2007shear, bridle2009handbook, refregier2014way, bruderer2015calibrated}. In all these examples the likelihood function is not tractable due to the highly non-gaussian nature of the signal, the fact that the data is in the form of a catalogue or maps and that the model and measurement process
is non-linear. 

In recent years, a new technique, known as Approximate Bayesian Computation (ABC), has gained attention in various fields such as population genetics, computational biology, ecology and psychology \cite{pritchard1999population, beaumont2002approximate, marjoram2003markov, sisson2007sequential, toni2009approximate, beaumont2009adaptive, mckinley2009inference, beaumont2010approximate, liepe2014framework, turner2012tutorial}. This method uses simulated data sets to bypass the need to evaluate a likelihood function. ABC systematically explores the prior model parameter space and compares the simulated and observed data sets using a distance metric. By accepting samples for which this distance metric is smaller than a given threshold, the method provides an approximation to the Bayesian posterior distribution. In addition, a summary statistic can be used to further compress the information in the data. Different algorithms have been proposed for performing these calculations efficiently. These include Importance Sampling, Markov Chain Monte Carlo, and different Sequential Monte Carlo methods \cite{beaumont2002approximate, pritchard1999population, marjoram2003markov, sisson2007sequential, toni2009approximate, beaumont2009adaptive, del2012adaptive}.

ABC methods have also started to be applied to problems in astronomy. For instance, Sequential Monte Carlo algorithms have been used for the model analysis of morphological transformation of galaxies \cite{cameron2012approximate}, the estimation of the luminosity function \cite{schafer2012likelihood} and the inference of cosmological parameters using TYPE Ia supernovae \cite{weyant2013likelihood}. Furthermore, a Markov Chain Monte Carlo variant of ABC was used to constrain the disk formation of the milky way \cite{robin2014constraining}. Simultaneously to this work, \cite{ishida2015cosmoabc} has published a software framework for likelihood-free inference based on ABC under the Cosmostatistics Initiative and \cite{lin2015new} applied the ABC scheme to predict weak-lensing peak counts.

In this paper, we explore the use of ABC for forward modeling in cosmology. After reviewing the principles of ABC, we consider its implementation using a Population Monte-Carlo algorithm. We test the resulting implementation using a Gaussian toy model. We then present an application of ABC to the calibration of wide field image simulations. We do this by calibrating image simulations generated by the \ufig software (Ultra Fast Image Generator) \cite{berge2013ultra} making use of the Mahalanobis distance metric \cite{mahalanobis1936generalized}. This represents a refinement of Monte Carlo Control Loops \cite{refregier2014way}, a forward modeling calibration method for weak lensing and other cosmological probes. The calibration finds a parameter configuration of \ufig that minimizes the difference between the simulations and the reference image using their statistical properties. This is important since a number of current high precision cosmological probes need to be tested for robustness against possible sources of systematics. These systematic errors can be numerous in nature and can also couple to each other in complex non-linear ways that are only evident when a realistic measurement is attempted. For these purposes creating simulations that have the same statistical properties as the data becomes crucial.

This paper is organized as follows. In section \ref{sec:framework}, we discuss the principles of Approximate Bayesian Computation and a particle based ABC algorithm as well as important performance considerations. Section \ref{sec:toymodel} compares Bayesian with ABC analysis on a Gaussian toy model. In section \ref{sec:calibration} we introduce the problem of image simulation calibration, discuss a distance metric to compare multivariate data sets and show results. Our conclusions are summarised in section \ref{sec:discussion}. We release a Python implementation of the ABC Population Monte-Carlo algorithm under GPLv3 license. Further details can be found in the Appendix \ref{sec:distribution}.

\section{Approximate Bayesian Computation}

Let us consider a data set $y$ and a model parametrised by a set of parameters $\theta$. From Bayes' theorem, the posterior probability of the model given the data is 
\begin{equation}
p(\theta|y) = \frac{p(y|\theta) p(\theta)}{p(y)},
\end{equation}
where $p(y|\theta)$ is the likelihood probability of the data given the model, $p(\theta)$ is the prior probability of the model and the normalisation comes from $p(y)$, the evidence. This expression can be used to derive the posterior from the likelihood and the prior.

\subsection{Principles}
\label{sec:framework}

Standard Bayesian inference relies on the evaluation of a likelihood. However, such a function is often not available for simulation-based models. In Approximate Bayesian Computation this problem is bypassed by considering a distance metric $\rho(x, y)$ that quantifies the difference between a simulated ($x$) and an observed ($y$) dataset. ABC algorithms sample the prior, $p(\theta)$, and a candidate parameter $\theta^*$ is accepted and retained as sample of the approximated posterior, if the distance $\rho(x, y)$ between $x$ and $y$ is less than a specified threshold $\epsilon$ \cite{beaumont2002approximate, pritchard1999population}. For small values of $\epsilon$, the ABC approximation to the posterior $p(\theta|y)$ is 

\begin{equation}
\label{eq:approx_posterior}
	p(\theta|y) \simeq p(\theta|\rho(x, y) \leq \epsilon),
\end{equation}

For complex data, it can be difficult or computationally expensive to calculate the distance $\rho(x, y)$ using all the information available in $x$ and $y$. Therefore, it is often useful to focus on summary statistics, $S(x)$ and $S(y)$, that capture the important features of the data such as the means and standard deviations. If a summary statistic contains the same amount of information about the model parameters as the whole data set, i.e. if  $p(\theta|y) = p(\theta|S(y))$,  it is referred to as being a sufficient statistic. If not it will lead to weaker posterior constraints. The ABC approximation to the posterior then becomes
\begin{equation}
\label{eq:approx_posterior_sum}
	p(\theta|y) \simeq p(\theta|\rho(S(x), S(y)) \leq \epsilon).
\end{equation}

Sequential importance sampling (SIS) \cite{beaumont2002approximate, pritchard1999population} is among one of the first algorithms proposed for ABC calculations. This method works by exploring the prior in parameter space and discarding all proposed points that do not fulfill the criterion $\rho(x, y) \leq \epsilon$. For small values of $\epsilon$, this method can become inefficient since the rejection rate can get very high. In 2003, ABC methods based on Markov Chain Monte Carlo (MCMC) were proposed \cite{marjoram2003markov}, which improved the sampling efficiency. More recently algorithms using Sequential Monte Carlo (SMC) with particle filtering \cite{sisson2007sequential, toni2009approximate, beaumont2009adaptive, del2012adaptive}  and \cite[for a review and tutorial see][]{turner2012tutorial}  have gained growing attention. 

Advanced algorithms such as ABC Population Monte Carlo (ABC PMC) are based on Sequential Monte Carlo (SMC) methods and try to circumvent the sampling inefficiency by constructing a series of intermediate distributions. These start from the prior distribution and converge to the approximate posterior (eq. \ref{eq:approx_posterior}) by sampling an intermediate distribution for a sequence of gradually decreasing thresholds $\epsilon_t$. Besides the mentioned ABC methods further algorithms and alternative likelihood-free frameworks exist (see \cite{turner2012tutorial} and references therein). 

In the following, we adopt the ABC PMC algorithm, as it requires only a small number of user supplied tuning parameters. We now discuss the details of this algorithm.

\subsection{ABC PMC algorithm}
\label{abcpmc}

The ABC PMC algorithm works with a large set of solution candidates, referred to as a `pool' in the following. Each candidate $\theta^*_i$ represents a position in parameter space and is referred to as a `particle'. The algorithm first generates an initial pool of $N$ particles, typically by randomly sampling from the prior $p(\theta)$, until all candidates fulfil the criteria $\rho(x, y) \leq \epsilon_0$, where $\epsilon_0$ is an initial threshold. Each particle in the pool is then assigned an initial weight $\omega_{i}=\frac{1}{N}$. In subsequent iterations, the algorithm moves to more stringent thresholds, resamples using the pool and updates the weights so that the particles sample the desired approximate posterior (eq. \ref{eq:approx_posterior_sum}).

The algorithm randomly samples from the pool taking into account the probability of each particle given by the assigned weight. Each sampled particle $\theta^*_i$ is then perturbed by randomly drawing from a Gaussian distribution with mean $\theta^*_i$ and a covariance matrix $\Sigma$ estimated from the particle positions of the current pool. The new particle $\theta^{**}_i$ is used to simulate a data set $x$ and if the data set passes the criterion $\rho(x, y) \leq \epsilon_t$ the particle $\theta^{**}_i$ is accepted ($\epsilon_t$ is the threshold chosen for iteration $t$). If particles are rejected, the process is repeated so as to maintain a constant population size of the pool. Finally, new weights are assigned to all the particles, which then determines the probability of being drawn in the next iteration. 

The weights allow the algorithm to favor particles from regions with high-probability and to reject particles from low-probability regions of the parameter space. The choice of weight $\omega_{i}$ has an important impact on the efficiency of the algorithm. Originally, proposed by \cite{beaumont2009adaptive}, the weight, $\omega_{i,t}$, in ABC PMC for particle $\theta_{i,t}$ at iteration $t$ is defined as 

\begin{equation}
\label{eq:weights}
	\omega_{i,t} = \frac{p(\theta_{i, t})}{\sum_{j=0}^{N} \omega_{j, t-1}q(\theta_{j, t-1}|\theta_{i,t},\Sigma_t)}
\end{equation}

where $p(\theta_{i, t})$ is the prior evaluated at position $\theta_{i,t}$. $\omega_{j,t-1}$ and $\theta_{j,t-1}$ are the weights and the particle position from the previous iteration, respectively. Finally, $q(\cdot|\theta_{i,t},\Sigma_t)$ is a Gaussian kernel with mean $\theta_{i,t}$ and covariance matrix $\Sigma_t$. The matrix $\Sigma_t$ is usually defined as twice the weighted variance of the particles $\theta_{i, t}$'s. Alternatives to eq. \ref{eq:weights} exist, such as replacing the Gaussian kernel with a multivariate Student-t distribution \cite{wraith2009estimation} or using an adaptive weighting scheme \cite{bonassi2015sequential}.

Algorithm~\ref{alg:abcpmc} gives a schematic view of the ABC PMC algorithm. A Python package containing an implementation of the algorithm is made publicly available under GPLv3 license. Its installation, instructions and documentation are described in detail in Appendix \ref{sec:distribution}.

\begin{algorithm}[h]
 \KwData{y, tolerance thresholds $\epsilon_t$, prior distribution $p(\theta)$}
 Set $t = 0$\;
 \For{$i = 0$ \KwTo $N$}{
 	\While{$\rho(x,y) > \epsilon_t$}{
 		Sample $\theta^{*}$ from the prior: $\theta^{*} \sim p(\theta)$\;
 		Create dataset $x$ from $\theta^{*}$ : $x \sim Model(\theta^{*})$\;
 	}
 	Set $\theta_{i, t} = \theta^{*}$\;
 	Set $\omega_{i, t} = \frac{1}{N}$\;
 }
 
 Set $\Sigma_t = 2 \times \Sigma(\theta_{0:N,t})$\;
 \For{$t = 1$ \KwTo $T$}{
 	\For{$i = 0$ \KwTo $N$}{
 		\While{$\rho(x,y) > \epsilon_t$}{
	 		Sample $\theta^{*}$ from the previous iteration: $\theta^{*} \sim \theta_{0:N,t-1}$ with weights $\omega_{i, t-1}$\;
	 		Perturb $\theta^{*}$: $\theta^{**} \sim \mathcal{N}(\theta^{*}, \Sigma_t)$\;
	 		Create dataset $x$ from $\theta^{**}$ : $x \sim Model(\theta^{**})$\;
 		}
	 	Set $\theta_{i, t} = \theta^{**}$\;
	 	Set $\omega_{i, t} = \frac{p(\theta_{i, t})}{\sum_{j=0}^{N} \omega_{j, t-1}q(\theta_{j, t-1}|\theta_{i,t},\Sigma_t)}$\;
 	}
 	Set $\Sigma_t = 2 \times \Sigma_{\omega_{0:N,t}}(\theta_{0:N,t})$\;
 }
 
 \caption{ABC Population Monte Carlo Algorithm (adapted from \cite{turner2012tutorial}). $\Sigma_{\omega_{0:N,t}}$ is the weighted empirical covariance.}
 \label{alg:abcpmc}
\end{algorithm}

Particle-based Sequential Monte Carlo algorithms offer advantages over commonly used MCMC algorithms. They are less likely to get stuck in low probability regions of parameter space and they reduce the difficulty of assessing the convergence of the sampling process. Both are especially true if the algorithms are used for ABC. Furthermore, the PMC algorithm can be trivially parallelized, which only advanced MCMC algorithms are suitable for \cite{foreman2013emcee, akeret2013cosmohammer}. Particle based SMC algorithms are easily parallelizable by construction, as the process of proposing and evaluating a new particle is independent of the remaining particle pool. Hence each particle could in principle be assigned to one CPU core.

Despite having shown to produce reliable results \cite[e.g.][]{sisson2007sequential, toni2009approximate, beaumont2009adaptive, mckinley2009inference, beaumont2010approximate, weyant2013likelihood, lin2015new}, it should be noted, however, that regular Bayesian inference should be favoured over the ABC method if an efficient likelihood evaluation is possible, as it usually offers better performance.

\subsection{Specific implementation}
\label{sec:performance}

For complex models the wall time is typically driven by the number of evaluations of the simulations. Therefore, having a high acceptance ratio is crucial in order to reduce the required time. The acceptance ratio in turn is driven by the choices of the thresholds as well as of the applied perturbation kernel. Using the appropriate balance in decreasing the threshold is important: if the decrease is slow, we expect a high acceptance ratio but on the other hand the true posterior is approximated only slowly. Decreasing the threshold fast results in a fast approximation of the posterior at the price of having a low acceptance ratio. 

Often the series of $\epsilon$ is manually selected, which can be difficult to define and may lead to a low acceptance ratio or a poor approximation of the posterior. Instead, an adaptive choice of the threshold is preferable. It has been proposed that the threshold $\epsilon_t$ should be set as the $\alpha^{th}$-percentile of the sorted particle distances $\rho(x,y)$ from the previous iteration \cite{beaumont2009adaptive, lenormand2013adaptive}, where $\alpha$ is a user defined value typically between $75$ and $90$. We find that this yields good results, both in terms of the acceptance ratio as well as of the final, approximated posterior. It has to be noted, however, that this approach can lead to a poor approximation to the posterior in certain cases \cite{silk2012optimizing}. 

The fact that the threshold is gradually adapted in the ABC PMC algorithm can be used to assess and monitor the convergence during the sampling process \cite{mckinley2009inference}. As $\epsilon_t$ approaches small values in practice, the approximated posterior (eq. \ref{eq:approx_posterior}) stabilizes and does not vary much in the subsequent iterations. A further reduction of the threshold $\epsilon_t$ does typically not improve the approximation significantly but will decrease the acceptance ratio \cite{mckinley2009inference, lin2015new}. In practical applications, a lower limit of the acceptance ratio has been applied as stopping criterion for the sampling \cite{lin2015new, ishida2015cosmoabc}. Further more theoretical convergence measurements have also been studied in \cite{barber2015rate}. Depending on the definition of the distance $\rho(x, y)$ the convergence of the sampling can be derived from the distribution of the distances. In particular if the expected variance of the stochastic model exceeds the threshold $\epsilon$ no further improvement of the approximation should be expected. Other convergence criteria have also been proposed, such as monitoring changes in the Kullback-Leibler (KL) divergence of the target densities \cite{wraith2009estimation} or thresholding on the so-called effective sample size (ESS) \cite{del2012adaptive, wraith2009estimation}.

The original algorithm proposed by \cite{beaumont2009adaptive} uses a Gaussian distribution with mean $\theta_{i,t}$ and twice the weighted covariance matrix of the particles as perturbation kernel. This choice minimizes the Kullback-Leibler distance between the desired posterior and the proposal distribution, which in turn maximizes the acceptance probability.

Lately, an alternative perturbation kernel has been proposed, which improves the acceptance ratio especially in non-linear, highly correlated parameter spaces \cite{filippi2013optimality}. The optimal local covariance matrix (OLCM) kernel is different for every particle $\theta_{i, t}$. It uses a multivariate normal distribution with a covariance matrix based on a subset of the particles from the previous iteration, whose distances are smaller than the threshold $\epsilon_t$ of the current iteration:

\begin{equation}
	\left\{ \left( \tilde{\theta}_{k, t}, \tilde{\omega}_{k,t} \right) \right\}_{0 \leq k < N_0} = \left\{ \left( \theta_{j, t}, \frac{\omega_{j,t}}{\bar{w}} \right) \text{ s.t. }  \rho(x, y) \leq \epsilon_t, 0 \leq j < N \right\} 
\end{equation}

\begin{equation}
\label{eq:cov}
	\Sigma_{\theta_{i, t}} \approx \sum_{k=0}^{N_{0}} \left[ \tilde{\omega}_{k}(\tilde{\theta}_{k} - \mu)(\tilde{\theta}_{k} - \mu)^T \right] + (\mu - \theta_{i, t})(\mu - \theta_{i, t})^T 
\end{equation}

where $\bar{w}$ is a normalization constant defined such that $\sum_{k=0}^{N_0} \tilde{\omega}_{k,t}=1$ and $\mu = \sum_{k=0}^{N_0} \tilde{\omega}_{k} \tilde{\theta}_{k}$. The covariance matrix $\Sigma_{\theta_{i, t}}$ in eq. \ref{eq:cov} is additionally corrected using a bias term to compensate the discrepancy between the mean of the particle population and the current particle $\theta_{i, t}$ (See \cite{filippi2013optimality}, Section 4.3.2 for detailed explanation on this kernel). Our experiments have shown that the OLCM kernel is able to increase the acceptance ratio while having a good exploration of the parameter space. 

\section{Gaussian toy model}
\label{sec:toymodel}

Let us consider the case where the data $y=\{y_1, y_2,.., y_n\}$ consists of independent and identically distributed (IID) samples $y_i$ drawn from a normal distribution with mean $\theta$ and  standard deviation $\sigma$. We will assume that the standard deviation $\sigma$ is known and will seek to evaluate the mean $\theta$ first using a Bayesian analysis and then comparing it with an ABC analysis.

\subsection{Bayesian analysis}

The probability distribution function (PDF) of a single sample $y_i$ given that the mean is $\theta$ is $P(y_i|\theta)=e^{-(y_i-\theta)^2/2 \sigma^2}/(\sigma\sqrt{2\pi})$. Since the variables $y_i$ are independent the likelihood is given by the joint probability
 
\begin{equation}
\label{eq:likelihood}
p(y|\theta) = \prod_{i=1}^{n} p(y_i|\theta) = \left( \sigma\sqrt{2\pi} \right)^{-n} \exp \left[-\sum_{i=1}^{n} \frac{(y_i-\theta)^2}{2 \sigma^2} \right]
\end{equation}

Assuming a flat prior (i.e. that $p(\theta) \propto$ constant), the normalised posterior probability is

\begin{equation}
\label{eq:posterior}
p(\theta|y) = \left( \frac{n}{2\pi \sigma^2} \right)^{\frac{1}{2}} \exp \left[ -\frac{n(\theta-\bar{y})^2}{2\sigma^2}  \right],
\end{equation}

where $\bar{y}=\sum_{i=1}^{n} y_i/n$ is the mean of the data points. Thus, in this simple case, the posterior probability distribution of the parameter $\theta$ is, a Gaussian with mean equal to the mean of the data and standard deviation $\sigma/\sqrt{n}$.

\subsection{ABC analysis}

Let us now pretend that we do not have the analytical expression for the likelihood (eq.~\ref{eq:likelihood}). We instead use ABC to estimate the posterior. For this purpose, we consider the average of the data points as a summary statistic
 
\begin{equation}
S(y)=\bar{y}
\end{equation}

We further consider the distance between the data $y$ and a simulated data set $x=(x_1,x_2,..,x_n)$ to be defined as

\begin{equation}
\label{eq:distance}
\rho[S(x),S(y)] = \left| S(x)-S(y) \right| = \left| \bar{x}-\bar{y} \right|
\end{equation}

For a given $\theta$, the average of the simulated data $\bar{x}$ is distributed like a Gaussian distribution with mean $\theta$ and standard deviation $\sigma/\sqrt{n}$. By considering the condition $\rho=\rho(x,y)<\epsilon$ and for a flat prior, one can then show that the ABC approximation to the posterior for a given threshold $\epsilon$ has the analytic form \cite{Wilkinson2008}

\begin{equation}
\label{eq:posterior_epsilon}
p(\theta|\rho<\epsilon) = \frac{1}{2\epsilon} \left[ \Phi\left( \frac{\bar{y}-\theta+\epsilon}{\sigma/\sqrt{n}} \right) - \Phi\left( \frac{\bar{y}-\theta-\epsilon}{\sigma/\sqrt{n}} \right) \right],
\end{equation}

where the cumulative distribution function (CDF) of the normal distribution can be expressed in terms of the error function as  $\Phi(t)=\left[ 1+ {\rm erf} (t/\sqrt{2}) \right]/2$. By Taylor expanding $\Phi$, it is simple to show that $p(\theta|\rho<\epsilon)$ is equal to the Bayesian posterior $p(\theta|y)$ (eq.~\ref{eq:posterior}) in the limit $\epsilon \rightarrow 0$. 

One can show that the expectation value of the posterior distribution is  $E[\theta|\rho<\epsilon]=\bar{y}$, while its variance is

\begin{equation}
\label{eq:abc_var}
{\rm var}[\theta|\rho<\epsilon]  = \frac{\sigma^2}{n} + \frac{\epsilon^2}{3}.
\end{equation}

Thus in the limit $\epsilon\rightarrow 0$ the ABC variance for $\theta$ (eq.~\ref{eq:abc_var}) reduces to the variance of the Bayesian posterior $\sigma^2/n$ (eq.~\ref{eq:posterior}). On the other hand as $\epsilon$ is increased, the ABC variance increases and so does the ABC acceptance rate, highlighting the trade off between the precision of the estimation of the parameter and the computational cost. Note that for large values of $\epsilon$ the variance diverges due to our improper flat prior.

\subsection{Results}

We use algorithm \ref{alg:abcpmc} with data $y=\{y_1, y_2,.., y_n\}$ defined as $n=10^4$ samples randomly drawn from a normal distribution with mean $\theta=1$ and standard deviation $\sigma=1$. As prior, we define $p(\theta) \propto$ constant for $-5 \leq \theta < 5$ and consider the distance $\rho(S(x), S(y))$ defined in eq. \ref{eq:distance}. To generate simulated data $x$ we use a normal distribution with mean $\theta^*$ and standard deviation $\sigma$. We set the initial threshold to $\epsilon_0=0.5$ and gradually decrease the threshold using $\alpha=90$ percentile of the particle pool. In each iteration, we create $N=2000$ particles and apply the multivariate normal kernel described in section \ref{abcpmc}.

\begin{figure}[t]
\begin{center}
\includegraphics[width=0.80\linewidth]{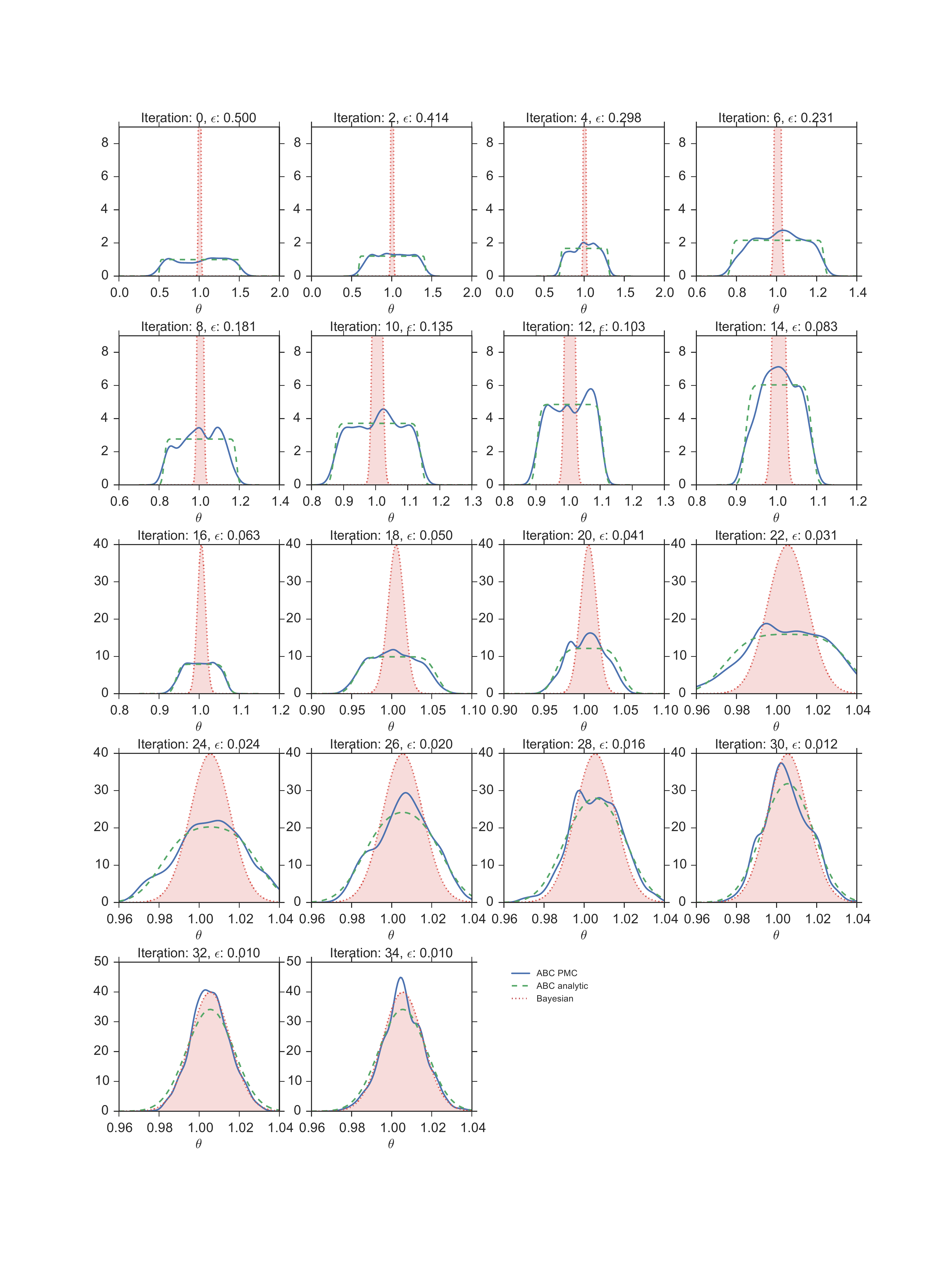}
\end{center}
\caption{Posterior distribution of the Gaussian toy model at different iterations $t$ and corresponding threshold values $\epsilon_t$. The numerical ABC PMC posterior is represented using a kernel density estimator (blue line) and is in good agreement with the analytical prediction (eq. \ref{eq:posterior_epsilon}, green line). As a comparison, the expectation for a Bayesian analysis is shown in red, which is the same function in all panels and was clipped when convenient.}

\label{fig:1d_gauss_posterior_evolution}
\end{figure}

\begin{figure}[t]
\begin{center}
\includegraphics[width=0.6\linewidth]{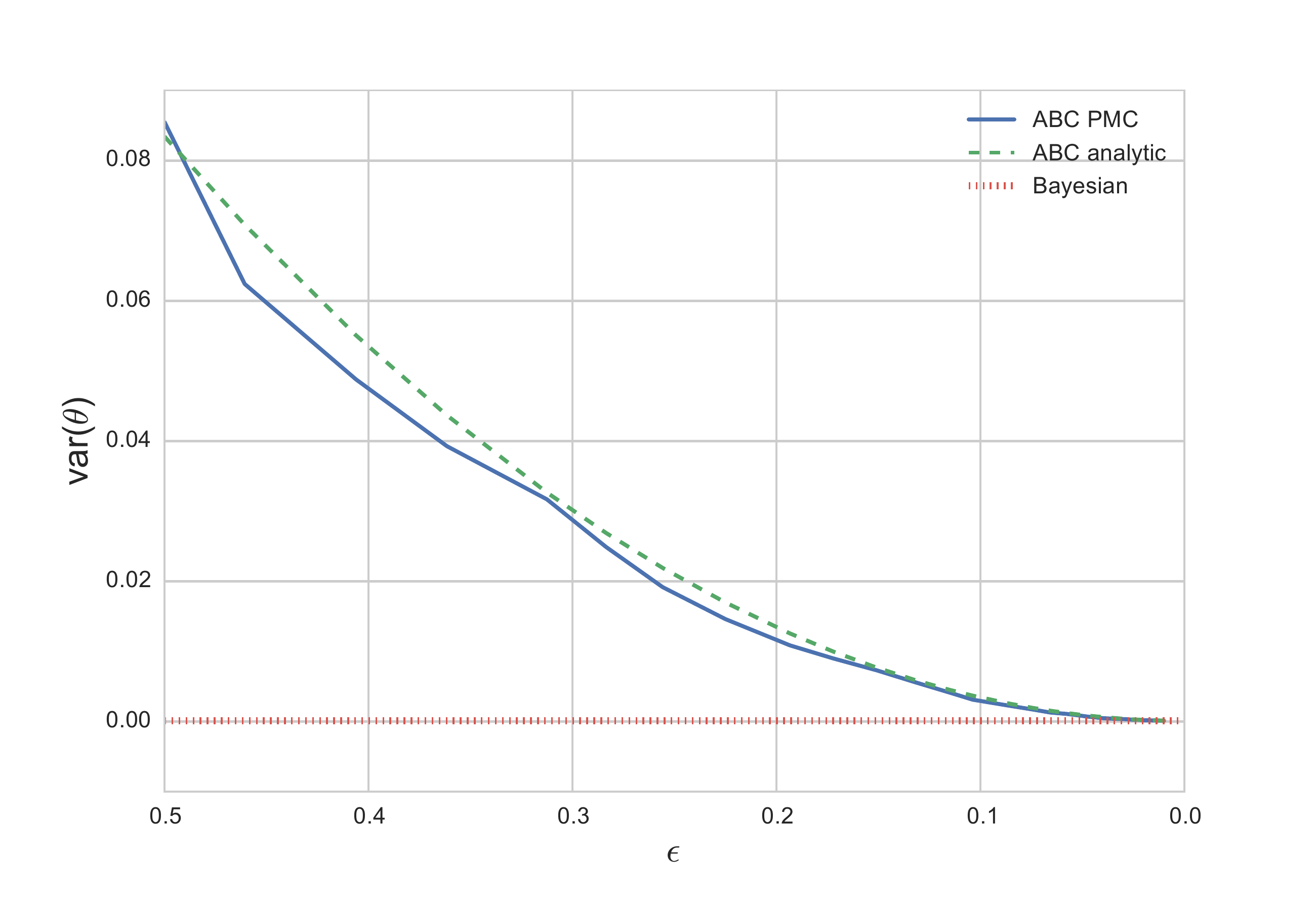}
\end{center}
\caption{Variance of $\theta$ as a function of the threshold values $\epsilon$ for the Gaussian toy model. The numerical ABC PMC variance (blue line) is in good agreement with the analytical prediction (green line) form eq. \ref{eq:abc_var} and with the Bayesian case (red line) as $\epsilon \rightarrow 0$. }
\label{fig:1d_gauss_variance}
\end{figure}

Figure \ref{fig:1d_gauss_posterior_evolution} shows the analytical posterior $p(\theta|\rho<\epsilon_t)$ (green line) given by eq. \ref{eq:posterior_epsilon} and a kernel density estimation (KDE) of the ABC posterior (blue line). Each panel depicts the results of different iterations $t$ with its corresponding decreasing threshold value $\epsilon_t$ (only even iterations are displayed). The sampled posteriors are in good agreement with the analytical prediction and for small values of $\epsilon_t$ they are close to the expected Bayesian distribution (red distribution in the figure). Further decrease of the threshold beyond 0.01 does not significantly change the posterior, which indicates the convergence of the estimation. This finding is supported by figure \ref{fig:1d_gauss_variance}, which shows the expected (eq. \ref{eq:abc_var}, green line), the approximated (blue line) and the Bayesian (red line) variance as a function of $\epsilon$. 

\section{Application to image modelling}
\label{sec:calibration}

The Ultra Fast Image Generator (\ufig) is a wide-field image simulation software package that generates realistic images and was optimised for fast computation \cite{berge2013ultra}. In \cite{bruderer2015calibrated} \ufig was used to generate image simulations with statistical properties consistent with observed images from the Dark Energy Survey (DES)  \cite{diehl2014dark}. This was done in the MCCL (Monte Carlo Control Loops) \cite{refregier2014way} framework by tuning the input parameters of the simulations. 

We now study how the ABC scheme can be applied to this problem, further details of which can be found in \cite{bruderer2015calibrated}.
Rather than comparing the images at pixel level, we first analyse them with the widely used \sextractor package \cite{bertin1996sextractor}, which produces a catalog of identified objects (e.g. stars and galaxies) and their properties (e.g size, flux and shape parameters).
These catalogs are used to construct summary statistics of the images and a distance metric between the simulations and a target image. The ABC iterative framework thus has the following steps:

\begin{description}

\item[Propose] The ABC PMC algorithm chooses and perturbs a particle from the pool and creates a new particle $\theta^*$ that represents a new position in parameter space.

\item[Create image] \ufig is parameterized using the proposed values in $\theta^*$ and generates a new image.

\item[Extract objects] \sextractor extracts the objects from the image, estimates their properties and generates a catalog.

\item[Postprocessing] The \sextractor catalog is post-processed to remove unphysical outliers and other artifacts and to compute derived properties such as object ellipticities. 

\item[Comparison] This catalog is then compared to the one from the target image using the distance metric.

\end{description}

This process is repeated for all ABC particles in each iteration while the threshold is gradually lowered until convergence is reached. 

\subsection{Distance metric}
\label{sec:distance}

In various practical applications, the impact of the choice of the distance metric on the approximated posterior have been studied. While a good choice plays an important role, sometimes even simple metrics have yielded good results \cite{lin2015new, weyant2013likelihood, mckinley2009inference, csillery2010approximate, beaumont2010approximate}. As summary statistics and their corresponding distance are typically problem specific, it is difficult to define metrics that are generally applicable. Instead domain knowledge and heuristics are being used in practice to define appropriate statistics. In \cite{blum2013comparative}, the authors reviewed techniques that can be used for the definition of summary statistics.

In the following we discuss distance metrics for \sextractor catalogs, where the statistical properties of multidimensional samples have to be compared. Quantifying the discrepancy between two multidimensional statistical distributions is non-trivial \cite[see e.g.][]{scott2009multivariate}. For univariate data sets, various statistical techniques have been developed to determine if two sets follow the same underlying PDF. A prominent example is the two-sample Kolmogorov-Smirnov test (KS test). Applying a KS test to multivariate data is not directly possible, especially beyond two dimensions. Applying the test to every dimension individually is typically insufficient, as correlations between different parameters are not taken into account. This is problematic for the image modeling application since, as we will see below, the object properties in the \sextractor catalogs are typically numerous and non-linearly correlated.

\begin{figure}[t]
\begin{center}
\includegraphics[width=0.6\linewidth]{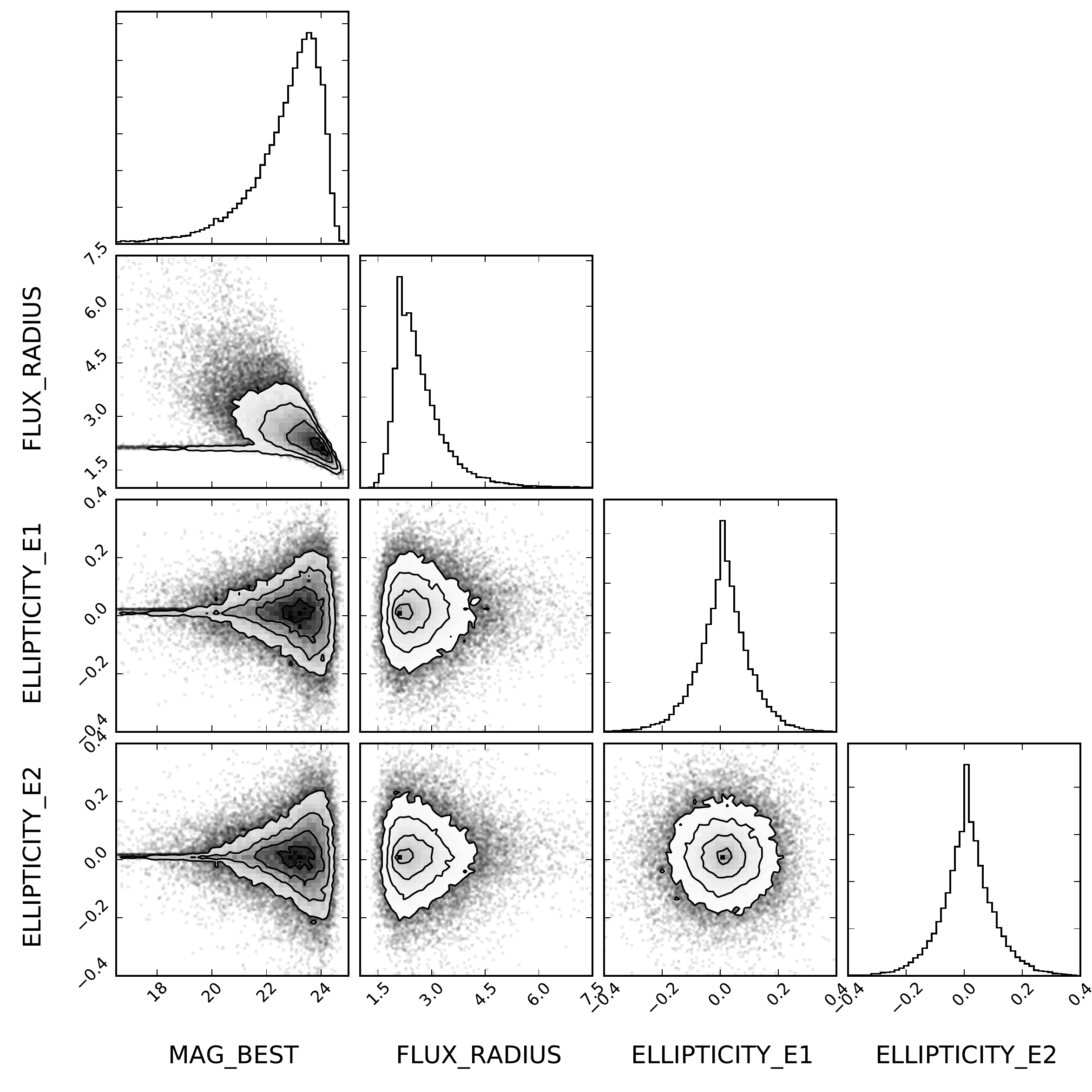}
\end{center}
\caption{Marginal distributions of the four selected colums of the \ufig \sextractor catalog from the target image. The plot shows the non-gaussianity and the non-linear correlations of the data set. Created with {\tt triangle.py} \cite{Foreman-Mackey:11020} }
\label{fig:sextractor_catalog}
\end{figure}

Diverse methods founded in information theory exist to quantify the difference between two multivariate distributions. These include the Kullback-Leibler divergences and its symmetrized variant the Jensen-Shannon divergence \cite{kullback1951information}. Both methods require the estimation of the underlying PDF. A common way to do this is to use a nearest neighbor or a kernel density estimator. However, both estimation methods tend to introduce an unwanted noise and bias in the distance measure \cite{budka2011accuracy}. Furthermore estimating the underling PDF is difficult in higher dimensions and is typically computationally intensive. Another approach is to define a distance metric between two multivariate data sets based on the Mahalanobis distance \cite{mahalanobis1936generalized}. We find that the Mahalanobis distance approach provides better constraints on the posterior while being computationally less demanding. For this reasons we opt for the latter in the following. 

The Mahalanobis distance between data vector $y$  (in our case derived from the target image) and a simulated data vector $x$ (from proposed simulated image) is based on the summary statistic,

\begin{equation}
\label{eq:mahalanobis}
	S(y)=\sqrt{(y - \mu_{y})^T\Sigma^{-1}_{y}(y-\mu_{y})}
\end{equation}
and 
\begin{equation}
\label{eq:maha_x}
	S(x)=\sqrt{(x - \mu_{y})^T\Sigma^{-1}_{y}(x-\mu_{y})},
\end{equation}

where $\mu_{y}$ is the mean of $y$ and $\Sigma_{y}$ its covariance matrix. Note that in eq. \ref{eq:maha_x} the data $x$ is compared to the center $\mu_{y}$ and covariance matrix $\Sigma_{y}$ of the observed data set $y$. As $S(x)$ and $S(y)$ are one-dimensional projections of $x$ and $y$ the distance $\rho(S(x), S(y))$ can be set to the standard one dimensional KS test for two-samples. In other words, the projection is the distribution of the distances to the center of the multidimensional distribution of the observed data set while taking into account the correlation in the data sets. The distance between the two projections is the maximal difference between the cumulative distribution functions (CDF) of their Mahalanobis distances.

As the Mahalanobis distance (eq. \ref{eq:mahalanobis} and \ref{eq:maha_x}) relies on the mean and the covariance matrix it is important to note that the proposed distance $\rho(S(x), S(y))$ works best on unimodal distributions. For data sets that heavily differ from a Gaussian, such as multimodal distribution for example, the distance might introduce an unwanted bias and yield improper approximations.

\subsection{Results}

In this section, we combine the ABC PMC algorithm described in section \ref{abcpmc} and the Mahalanobis distance metric to constrain the \ufig simulation parameters to mimic a given target image. 

For this purpose we generated a target image using the parameters $\hat{\theta}$ shown in the second column of table \ref{tab:parameters} \cite[see][for details]{berge2013ultra, bruderer2015calibrated}:

\begin{itemize}
  \item {\tt size-sigma} defines the root mean square of the size (r50) distribution of galaxies in arcsec,

  \item {\tt size-theta} is the correlation angle for size-magnitude distribution of galaxies,
    
  \item {\tt e1-sigma} and {\tt e2-sigma} are root mean square of the two components of the galaxy ellipticities e1 and  e2.
\end{itemize}

For the example explored here all the other simulation parameters are kept fixed to values similar to those in \cite{berge2013ultra}. We run the ABC PMC algorithm \ref{alg:abcpmc} with the OLCM permutation kernel from Section \ref{sec:performance} and $N=400$ particles on these four simulation parameters. The initial threshold $\epsilon_0$ was set to $0.2$ and automatically reduced by using $\alpha=90$ percentile. To calculate the distance metric we used the following \sextractor catalog columns:

\begin{itemize}
  \item {\tt MAG-BEST} is the estimated magnitude of the objects, 
  
  \item {\tt FLUX-RADIUS} is the half light radius of an object,  
  
  \item {\tt ELLIPTICITY-E1} and {\tt ELLIPTICITY-E2} are the two components of ellipticities of the objects. 
\end{itemize}

The one- and two-dimensional marginal distributions of these four catalog properties from the target image, are shown in figure \ref{fig:sextractor_catalog}. As stated earlier, these are highly and non-linearly correlated. As priors, we use a component-wise Gaussian distributions with means and standard deviations shown in the third column of table \ref{tab:parameters}. The widths of the priors are choosen such that they exeed the $95\%$ confidence limits shown in table 2 in the appendix of \cite{bruderer2015calibrated}. Furthermore, the means of the priors have been defined to be shifted by at least $1\sigma$ from the true target value

The calculation was parallelized on 200 Intel Ivy Bridge EP E5-2660v2 2.2 GHz cores and resulted in a wall time of approximately 20 hours.

\begin{figure}[t]
\begin{center}
\includegraphics[width=0.8\linewidth]{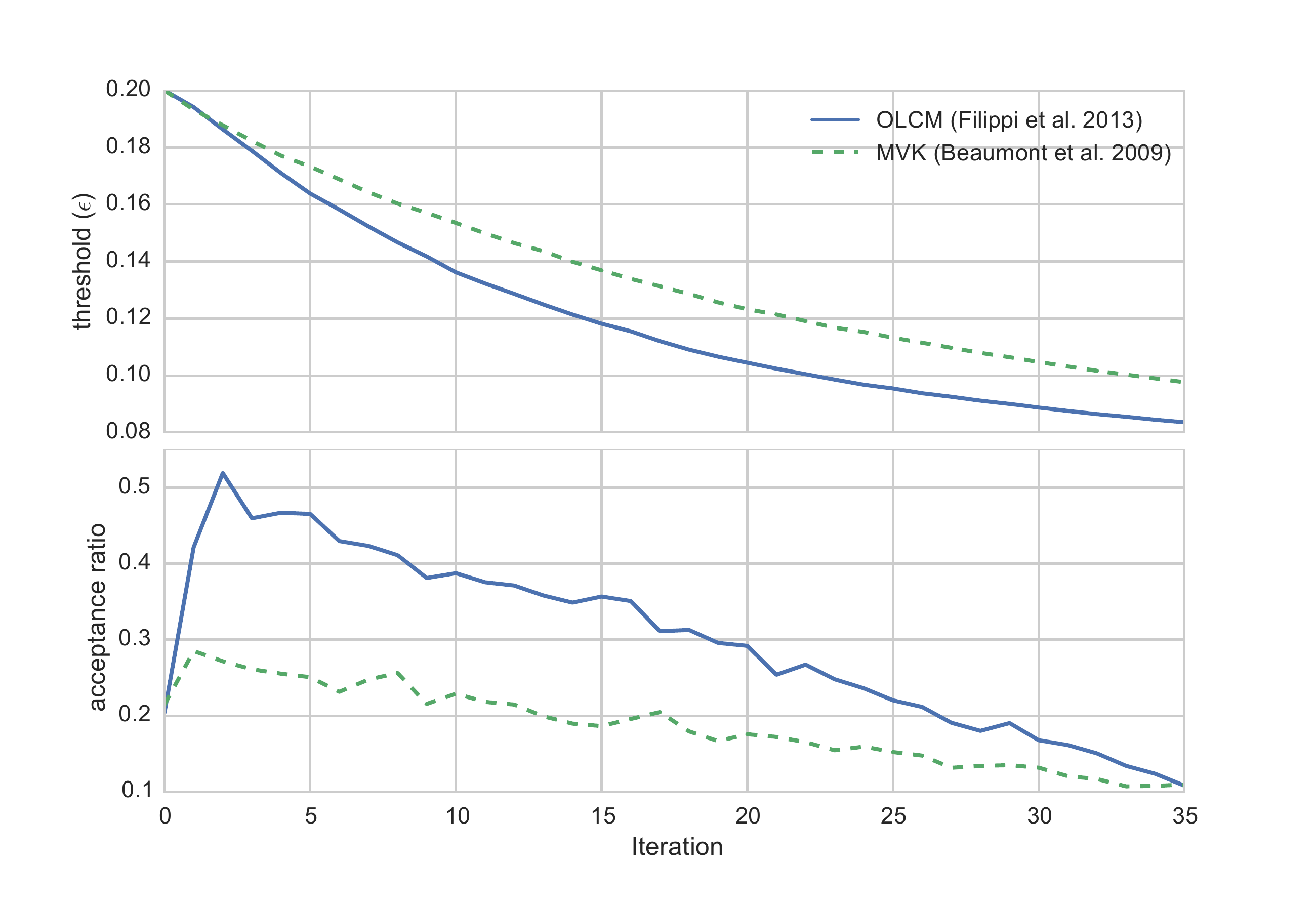}
\end{center}
\caption{The threshold $\epsilon$ and the acceptance ratio as a function of the iterations for the \ufig image modeling. The blue line shows the multivariate kernel (MVK)\cite{beaumont2009adaptive} and the green line denotes the optimal local covariance matrix kernel (OLCM) \cite{filippi2013optimality}.}
\label{fig:eps_acc}
\end{figure}

\begin{figure}[t]
\begin{center}
\includegraphics[width=0.7\linewidth]{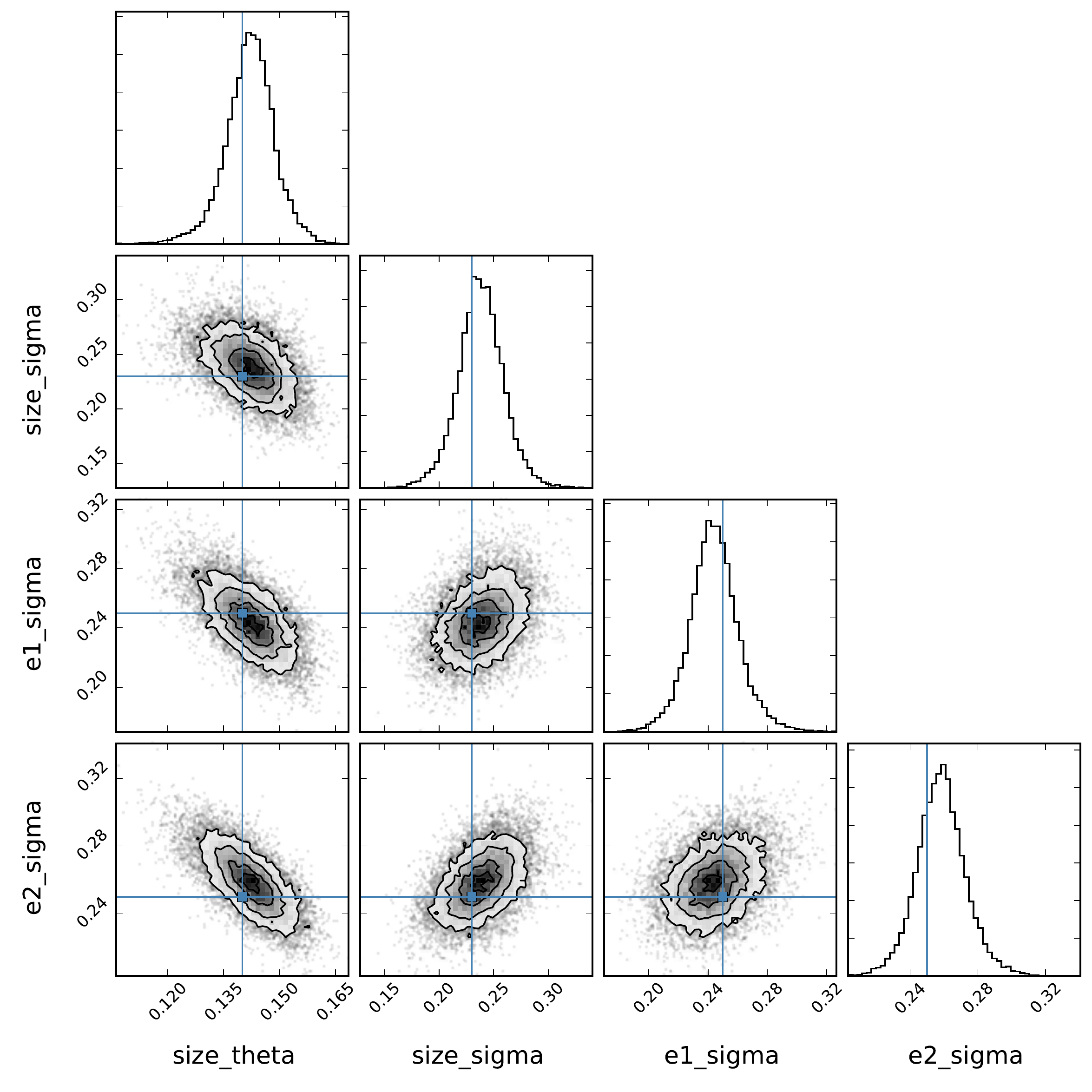}
\end{center}
\caption{The one- and two-dimensional marginal distributions of the approximate \ufig parameter posterior. The blue lines denote the true initial parameter configuration. Created with {\tt triangle.py} \cite{Foreman-Mackey:11020}}
\label{fig:ufig_posterior}
\end{figure}

Figure \ref{fig:ufig_posterior} shows the marginal distributions of the ABC posterior. The blue lines denote the true parameter values used to generate the target image, which are consistent with the approximated posterior. The approximated posterior is well behaved and displays correlation between different parameters. The means and standard deviations of the parameters estimated from the approximate posterior are shown in the fourth column of table \ref{tab:parameters}. The ABC PMC algorithm was thus able to refine the prior information and correctly moved the mean towards the target parameter values and reduced the errors on all parameters. The estimated errors are in grood agreement with the $95\%$ confindence limit in \cite{bruderer2015calibrated} Figure \ref{fig:eps_acc} shows the behavior of the thresholds $\epsilon_t$ and the acceptance ratio as a function of the number of iterations for the normal multivariate and the OLCM kernel. The OLCM kernel allows for a faster decrease of the threshold while having a higher acceptance ratio.

\begin{table}[tdp]
\caption{Target image parameter configuration, prior definition and ABC posterior. The uncertainties corresponds to one standard deviation.}
\begin{center}
\begin{tabular}{l|r|r|r}
Parameter & \pbox{20cm}{Target } & \pbox{20cm}{Prior} & \pbox{20cm}{ABC } \\
\hline
\hline

{\tt size-theta} & 0.14 & 0.15$\pm0.03$ & 0.1420$\pm0.0047$\\
{\tt size-sigma} & 0.23 & 0.21$\pm0.07$ & 0.2376$\pm0.0154$\\
{\tt e1-sigma}   & 0.25 & 0.26$\pm0.07$ & 0.2435$\pm0.0119$\\
{\tt e2-sigma}   & 0.25 & 0.24$\pm0.07$ & 0.2585$\pm0.0104$\\

\end{tabular}
\end{center}
\label{tab:parameters}
\end{table}%

\section{Conclusion}
\label{sec:discussion}

We explored how Approximate Bayesian Computation can be applied to forward modeling in cosmology, where the likelihood is unavailable or intractable. We discussed a common implementation of the ABC algorithm, the Population Monte Carlo (PMC) algorithm, which is a combination of Sequential Monte Carlo and particle filtering. The algorithm performs the approximation of the posterior distribution by using a pool of particles that are within a certain distance threshold. The solution is iteratively improved by gradually lowering this threshold value. We discuss several considerations for good performance of the algorithm such as an automatic reduction of the acceptance threshold and the impact of particle permutation kernels. 

We apply the ABC PMC algorithm to a Gaussian toy model as well as to the calibration of an image generated with the simulation software (\ufig). We show that the analytical predictions for the toy model are in good agreement with our empirical results and approach the Bayesian posterior in the limit of small thresholds. For the image calibration application, we introduce a distance measure based on a parameter space projection using the Mahalanobis distance that can be used to measure the discrepancy between two multivariate distributions. To assess the goodness of the inferred solution, we have generated an image with known configuration and compared the estimated posterior with the input configuration. We find that ABC produced a reliable approximate posterior that was consistent with the input parameter values and mapped the correlation between simulation parameters. The ABC method with its PMC implementation is thus promising for numerous forward modeling problems in cosmology and astrophysics. 

\section*{Acknowledgements}

We would like to thank Sarah Grimm of the statistical consulting group of ETH Zurich for the support in the analysis of distance measures. We also thank Claudio Bruderer, Joel Berge and Lukas Gamper for useful discussions during the development of the image model calibration software. This work was supported in part by grant 200021\_143906 from the Swiss National Fundation.

\appendix

\section{Package distribution}
\label{sec:distribution}

Detailed documentation, examples and installation instructions for the ABC PMC implementation can be found on the package website \url{http://abcpmc.readthedocs.org/}. The package is released under the GPLv3 license and has been uploaded to PyPI\footnote{\url{https://pypi.python.org/pypi/abcpmc}} and can be installed using pip\footnote{\url{www.pip-installer.org/}}: 

\begin{verbatim}
$ pip install abcpmc --user
\end{verbatim}

This will install the package and all of the required dependencies. The development is coordinated on GitHub \url{http://github.com/jakeret/abcpmc} and contributions are welcome.

The package is entirely written in Python and contains the algorithm \ref{alg:abcpmc} as well as various threshold schemes, prior implementations and different pertubation kernels. The code is built with a flexible design such that one can easily extend the provided functionality.

\bibliographystyle{elsarticle-num}

\bibliography{ufigcalib}

\end{document}